%% file: main.tex
\documentclass[sigconf]{acmart}
\usepackage{amsmath,color,graphicx}

\AtBeginDocument{%
  \providecommand\BibTeX{{%
    \normalfont B\kern-0.5em{\scshape i\kern-0.25em b}\kern-0.8em\TeX}}}

\setcopyright{acmcopyright}
\copyrightyear{2018}
\acmYear{2018}
\acmDOI{10.1145/1122445.1122456}

\acmConference[Woodstock '18]{Woodstock '18: ACM Symposium on Neural
  Gaze Detection}{June 03--05, 2018}{Woodstock, NY}
\acmBooktitle{Woodstock '18: ACM Symposium on Neural Gaze Detection,
  June 03--05, 2018, Woodstock, NY}
\acmPrice{15.00}
\acmISBN{978-1-4503-9999-9/18/06}

\begin{document}

\title{CQBG* : Advanced Academic Team Worker Recommendation Models}

\author{Mi Wu}




\begin{abstract}

\input{abs.tex}
\end{abstract}



\maketitle

\input{intro.tex}

\input{relatedwork.tex}

\input{problemdef.tex}

\input{method.tex}
\input{experiments.tex}

\input{conclusion.tex}

\balance

\bibliographystyle{ACM-Reference-Format}
\bibliography{ref}
\end{document}

%% file: abs.tex
Collaborator recommendation is an important task in academic domain. Most of the existing approaches have the assumption that the recommendation system only need to recommend a specific researcher for the task. However, academic successes can be owed to productive collaboration of a whole academic team. In this work, we propose a new task: academic team worker recommendation: with a given status: student, assistant professor or prime professor, research interests and specific task, we can recommend an academic team formed as (prime professor, assistant professor, student). For this task, we propose a model CQBG-R(Citation-Query Blended Graph-Ranking). The key ideas is to combine the context of the query and the papers with the graph topology to form a new graph(CQBG), which can target at the research interests and the specific research task for this time. The experiment results show the effectiveness of the proposed method. 

%% file: intro.tex
  \section{Introduction}
  Collaborator recommendation is an important topic in information retrieval and data mining community. 
 It has many application in real world. For example, in technology companies, the program manager may want to form a team to development new produces; in sport competition, a coach may want to find some new team members to strength their ability; in academic community, some researchers may want to form a team to publish some high quality papers. 
 
 A lot of works have been done in the academic team recommendation area, these methods usually includes finding experts in a specific area, recommending coauthor for a researcher or finding advisor for an applying student. Despite all the great achievements,  there still have some disadvantages. One of the disadvantage of these methods is all the collaborator recommendation is from one \textbf{Person}  perspective. However, nowadays' academic achievements like interdisciplinary projects or papers often need the efforts of many people in the form of a team. Different people take different roles in the team to extend the boundary of the research. For example, the prime professor needs to point out the direction of the whole project clearly, young assistant professors must have a good understanding of the detail of the research and advise students and PhD students have to come up with some novel ideas and implement the algorithms. In this scenario, recommending collaborators from one \textbf{Person} perspective is on a coarse resolution and makes no sense like recommending an old professor in high position for a first-year PhD student.

 In recent years, team formulation has been studied from different aspects like team member replacement and the performance of the whole team with regard to each member. In addition, the development in network representation learning provides effective tools for more challenging recommendation tasks. 
 So, for this research track project, we try academic team worker recommendation. For this task, given some conditions like field: data mining, information retrieval or machine learning, the academic level: student, assistant professor or prime professor and other information, we will recommend the remaining partners in an academic team for these conditions. 
 
 To solve this problem, we develped a models CQBG-R(Citation-Query Blended Graph-Ranking). 
 We first extract the text information from the DBLP dataset and build a citation network according to the collaboration between different researchers, and we call this network CQBG.
 The contribution can be formed as the follow:
  \begin{itemize}
      \item The first team recommendation work in academic research domain.
      \item We have combined the topology of the graph and the context information of queries and papers and brought up a novel graph-CQBG(Citation-Query Blended Graph-Ranking). 
      \item Our model can achieve good academic team worker recommendation performance and interpretability. 
  \end{itemize}

%% file: relatedwork.tex
\section{Related Work}
\subsection{Network Representation Learning}
Network is an important type of data storage. As deep learning has been applied successfully to image and language sequence, researchers try to introduce this powerful tool in to data in the form of network. However, compared to image or language sequence, network lacks a fixed kind of data form, which is easy to be handled. So, Network representation learning has attracted a lot of attention. \textbf{Network Representation Learning} also refers to \textbf{Network Embedding}, which means encoding the information of the whole network into low dimensional vector. Many algorithms of network representation learning have been raised in recent years. Theses algorithms can mainly be divided into two categories:
\begin{itemize}
    \item Algorithms for network without attributes
    \item Algorithms for network with attribute information
\end{itemize}
Network without attributes means the network $G$ is composed of $(V,E)$, $V$ is the set of nodes:
  \{$v_{1}$,$v_{2}$,$\cdots$, $v_{n}$\} and $E$ denotes the set of all edges: \{$e_{1}$,$e_{2}$,$\cdots$,$e_{m}$\}. all the information stored in network without attributes is reflected in the topology in the network like the connection between nodes and the degree of nodes. So, algorithms targeting at network without attributes try to encode the topology of network into the low dimensional vector. In this learning process, the vector should be able to reconstruct the whole network. For example, the nodes closed to each other should have similar vectors. Based on such assumption, researchers raise many models like DeepWalk\cite{Perozzi:2014:DOL:2623330.2623732}. Algorithms for network without attributes achieve good performance in some tasks. However, networks in real life can be much more complex such as some network's nodes may have content or attribute attached to them or the nodes and edges have various types. For these networks, algorithms like \cite{Ou:2015:NHL:2783258.2783283,Ou2016AsymmetricTP,wang2017signed} have been raised.
\subsection{Collaborator Recommendation}
 Recommendation systems have been studied for a long time. They range from online websites recommendation ~\cite{amazon2003}, mobile applications ~\cite{Xu2011imc} to academic recommendation ~\cite{Tang2012kdd}.The existing research work about collaborator recommendation can be partitioned into two main categories (Topic based model and non-topic based model). The topic based model could be divided into two subcategories: cross domain recommendation and single domain recommendation. And the non-topic based model could be divided into three subcategories: heterogeneous network based recommendation, homogeneous network based recommendation and non-graph based model. In a real academic network, there are multiple type of attributes among nodes, e.g. people, paper, conference, and there are multiple types of links among nodes, e.g. publish, write. Without considering these information, the performance wouldn't be great. sun et al. ~\cite{sun2011} proposed a Collaborator Recommendation algorithm on heterogeneous network, which aims at predicting whether two authors in the network who have never co-author before will co-author sometime in the future. In ~\cite{Sun11pathsim}, the authors recommends Collaborators in heterogeneous network according to the similarity between two researchers. They used the number of common meta-path between two researchers to measure the similarity between them. Different from the heterogeneous network based recommendation, it aims at recommending collaborators who work at a specific search topic. In other words, given a specific researcher $R$ and the topic $T$, e.g. data mining, machine learning, recommendation with Topics will find the candidates who are most likely to work with $R$ on the required topics. Liu et al. ~\cite{Liu2018kdd} proposed an algorithm to solve "Context-aware Academic Collaborator Recommendation". "Context-aware" means considering the topic information when recommend collaborators. Tang et al.~\cite{Tang2012kdd} proposed a cross domain collaboration recommendation with topic information of researches. Despite its popularity in both industry and research community, only a few of researches have been proposed for collaborator recommendation or team member recommendation, especially for academic domain. So, in our research , we will try recommending academic team for a specific target.

\subsection{Graph Reasoning}

Graph reasoning has been studied for a long time ~\cite{liu2022knowledge, Yan_Liu_Ban_Jing_Tong_2021, liu2023conversational, liu2023knowledge}, and it has many applications, such as knowledge graph completion ~\cite{xu2022abm}, question answering ~\cite{newlook, gfinder, binet, prefnet}, fact checking ~\cite{kompare, inspector, liu2022knowledge}, entity alignment ~\cite{Yan_Liu_Ban_Jing_Tong_2021}, recommender system ~\cite{recommendation, du2023neural} and so on. Under the umbrella of knowledge graph reasoning, conversational question answering is an important task ~\cite{liu2023conversational}.

%% file: problemdef.tex
\section{Problem Definition}

In this paper, we aim to recommend the top $k$ team members for a user with a specified research interest and a prefer topic.
For convenience expression, we also use task query to denote prefer topic in this paper.
We denote the name of the user as $N$, the research interest as $r$ and the prefer topic/task query as $q$. 
Here we give an example to show the goal of our task. Assume we want to recommend a team for professor Jiawei Han, and his research interest is Data Mining, his prefer topic is heterogeneous network. And he wants to form a team with three different roles which are $(Prime Professor, Assistant Professor, Student)$. Because Jiawei Han is a prime professor, so our algorithm will recommend two different role researchers to him. These researchers belongs to two different roles which are $Assistant Professor$ and $Student$. More specifically speaking, our algorithm will recommend "Yizhou Sun" and “Junheng Hao“ to Professor Jiawei Han who are $Assistant\ Professor$ and $PhD\ Student$ in University of California, Los Angeles, respectively.

\begin{definition}{Role} 
Role is the group of a user belongs to. e.g.Prime Professor, Assistant professor, Student.
\end{definition}

\begin{definition}{Research interest} 
Research interest is the research area of the user.
\end{definition}

\begin{definition}{Query} 
Query is the project topic the user like to work, e.g. "Heterogeneous network mining".
\end{definition}

\begin{definition}{Citation Graph} 
Citation graph is the graph of citations between users.
\end{definition}

\begin{definition}{Query Graph} 
Query graph is the graph got by BM25.
\end{definition}

\begin{definition}{CQBG} 
   CQBG is the graph merged by Citation Graph and Query Graph.
\end{definition}

\begin{table}[h]
	\centering
	\caption{Notations and Definition}
	\vspace{-0.5\baselineskip}
	\begin{tabular}{|c|l|}
		\hline
		Symbols       & Definition                \\ \hline
		$n_i$  & the node needed to be recommend\\ \hline
		$N$ & the set of all authors\\ \hline
		$G^l$=\{$N$, $E^l$\} & an attributed query graph \\ \hline
		$G^c$=\{$N$, $E_G$\} & an attributed data graph  \\ \hline
		$G = (N, E)$ & Citation-Query Blended Graph  \\ \hline
		$q$ & task query \\ \hline
		$F(n_j | n_i, q)$ & matching function       \\ \hline
		$e_i$ & the embedding vector of node $v_i$ \\ \hline
		$v_i$ & the node in the graph \\ \hline
	\end{tabular}
\label{table1}
\vspace{-0.8\baselineskip}
\end{table}

Given: A researcher name $n_i$, the role of the researcher, the research interest $r$, a task query $q$ and a user input parameter $k$

Output: The top $k$ team pairs, and each pair and $n_i$ form three different roles which are $(Prime Professor, Assistant Professor, Student)$.

%% file: method.tex

\section{Methodology}

Before the elaboration of our model, we first present the data we used and the information we utilized to construct the graph.
\smallskip

\subsection{Dataset}
\noindent \textbf{Citation Network} \footnote{http://arnetminer.org/citation} incorporates bibliographic information derived from the mainstream journals and proceedings of computer science. It is a list of items in which each item contains necessary information of one paper, including title, authors, abstract, publishing time, name of conference or journal, citation list (the papers that cited this paper). In detail, the dataset contains 2,244,021 papers and 4,354,534 citations.


\subsection{ Citation-Query Blended Graph }

First we divided the information of one paper into two classes, relation and text data. The relation data contains the author and citation list, contrary to the text data that consists of the title and abstract. To fully leverage these two types of data, we propose the Citation-Query Blended Graph.
\smallskip

\noindent \textbf{Citation Graph.} We build the co-authorship graph $G^c(N,E^c)$ based on the relation data, in which each node is denoted by an author and an edge represent the relation between two authors quantified by their past collaborated work. In the research community, for example, the edge weight between two scientists can be related to the number of publications they have collaborated. Here, we count the citations of each paper and sum the citations of papers of they collaborated, as the following.

Given two authors $n_i$ and $n_j$, the edge weight $e^c_{ij}$ of $n_i$ and $n_j$ is defined as:
\begin{displaymath}
e^c_{ij} = \sum_{p \in D(n_i, n_j)}  C(p),
\end{displaymath}
where $D(n_i, n_j)$ is the set of papers where $n_i$ and $n_j$ are collaborated and $C(p)$ is the number of citations of paper $p$.

\smallskip
\noindent \textbf{Query Graph.} Recall that the user will input a task query, which should account for how the works of an author are related to the query. There, we leverage the text data to measure the correlation between the author and the query. First, we make a document for each paper simply by incorporating the title and abstract, denoted by $D$. We use $D(n_i, n_j)$ to represent the papers collaborated by $n_i$ and $n_j$.  By modifying a famous model BM25\cite{bm25}, we assign an edge weight for any two authors, building query graph $G^q(N, E^q)$, as the followings.

Given a query $q$ and two authors $n_i$ and $n_j$, the edge weight $e^q_{ij}$ of $n_i$ and $n_j$ is defined as:
\begin{displaymath}
e^q_{ij} = \sum_{d \in D(n_i, n_j)}\text{BM}(q, d),
\end{displaymath}
and 
\begin{displaymath}
\text{BM25}(d(n_j), q) =  \sum_{w\in q}c(w, q)c'(w,d) \log\frac{|\mathbf{D}|+1}{\text{df}(w)},
\end{displaymath}
where 
\begin{displaymath}
c'(w,d) = \frac{(k+1)c(w,d)}{c(w,d)+k(1-b+b\frac{|d|}{\text{avg}(d)})}.
\end{displaymath}
$c(w, q)$ is the frequency of $w$ occurring in $q$ and $c(w,d)$ is the frequency of $w$ occurring in $d$. $\mathbf{D}$ is the set of all the documents and $\text{df}(w)$ is the number of papers having $w$.
\text{avg}(d) is the average length of all documents. 
$k$ and $b$ are two parameters to penalize the length of document and $c(w,d)$.

In this paper, we just use their default value $1.5$ and $0.75$. 

\noindent \textbf{CQB Graph.} With the citation graph and query graph, we blend then into a graph $G(N, E)$.

Given two node $n_i$ and $n_j$,  the edge weight $e_{ij}$ between $n_i$ and $n_j$ is computed as:
\begin{displaymath}
e_{ij} = \text{Norm}(e^c_{ij}) + \text{Norm}(e^l_{ij}) 
\end{displaymath}
where Norm() denotes a normalization function, 
\begin{displaymath}
\text{Norm}(e^c_{ij}) = \frac{e^c_{ij} - min\{e'^c_{ij}: e'^c_{ij} \in E^c\} }{max\{e'^c_{ij}: e'^c_{ij} \in E^c\} - min\{e'^c_{ij}: e'^c_{ij} \in E^c\} }
\end{displaymath}
and 
\begin{displaymath}
\text{Norm}(e^l_{ij}) = \frac{e^l_{ij} - min\{e'^l_{ij}: e'^l_{ij} \in E^l\} }{max\{e'^l_{ij}: e'^l_{ij} \in E^l\} - min\{e'^l_{ij}: e'^l_{ij} \in E^l\} }
\end{displaymath}
Therefore $e_{ij}$ combines two scores out of which one is to measure the work effectiveness of two users and author one is to capture the correlation between the their collaborated works and the query.

Next, we will design a traditional model and a graph representation learning model.

\subsection{Hierarchical Tree}
Given a worker $n_i$ who needs to be recommended an academic team, we need design a model to rank all the remaining authors. Note that in this paper, we totally assume three roles, which indicates we need recommend other two role for $n_i$. Therefore, we provide three criteria to estimate the academic level for each author.

\noindent \textbf{Criteria 1: Paper}. There, we need two thresholds $t_1$ and $t_2$, $t_1< t_2$. Let $|D(n_i)|$ be the number of papers $n_i$ published. Then 
$$ n_i=\left\{
\begin{aligned}
&\text{prime professor} \ \  &|D(n_i)| > t_2 \\
&\text{assistant professor}  \ \ & t_2 \geq |D(n_i)| > t_1 \\
&\text{student}   \ \   &|D(n_i)| \leq t_1 .
\end{aligned}
\right.
$$
\noindent \textbf{Criteria 2: Citation}. Given two thresholds $t_1$ and $t_2$, let $D(n_i)$ be the number of papers $n_i$ published, and $C(p)$ be the number of citations of this paper, $\mathbf{C}(n_i) = \sum_{p \in D(n_i)}C(p)$   . Then 
$$ n_i=\left\{
\begin{aligned}
&\text{prime professor} \ \  &\mathbf{C}(n_i) > t_2 \\
&\text{assistant professor}  \ \ & t_2 \geq \mathbf{C}(n_i)> t_1 \\
&\text{student}   \ \   &\mathbf{C}(n_i) \leq t_1 .
\end{aligned}
\right.
$$

\noindent \textbf{Criteria 3: Neighbor}. Given two thresholds $t_1$ and $t_2$, let Degree$(n_i)$ be the degree of $n_i$ in $G$, which indicates the number of authors $n_i$ has collaborated with. Then

$$ n_i=\left\{
\begin{aligned}
&\text{prime professor} \ \  &\text{Degree}(n_i)  > t_2 \\
&\text{assistant professor}  \ \ & t_2 \geq \text{Degree}(n_i) > t_1 \\
&\text{student}   \ \   &\text{Degree}(n_i) \leq t_1 .
\end{aligned}
\right.
$$

In this paper, we used the Criteria 1 to classify all the nodes and $t_2$ is set as $40$ and $t_1$ is set as $20$.

\subsection{CQBG-R}
Given a worker $n_i$ who needs to be recommended an academic team, we design  a model to rank each class nodes respectively. Note that if we know the role of $n_i$, we only rank other two classes. However, if we do not its role, we will assign a role for $n_i$ estimated by the criteria mentioned above. 

Let $n_j$ be a node of a class needed to be scored. Then we design our first model based on two intuitions:
\begin{enumerate}
    \item the recommended author should have good reputation and correlation with the query based on his/her past collaborations with other authors.
    \item the recommended author should be closed to $n_j$ in order to have low communication cost.
\end{enumerate}

We quantify the first intuition by computing the average weight of $n_j$, because each edge contains the information of their reputation of collaborated works (e.g., citations) and the correlation of their works with query (e.g., BM25 score), as the followings:
\begin{displaymath}
I_1 = \frac{\sum_{j' \in N(n_j) } e_{ij}}{ |N(n_j)|}
\end{displaymath}
where $N(n_j)$ is the set of neighbors of $n_j$. 

To measure the intuition2, we utilize the shortest path in $G$, i.e., they have lower communication cost if they have shorter path, as the followings:
\begin{displaymath}
I_2 = \frac{1}{|SP(n_i, n_j)|}
\end{displaymath}
where $|SP(n_i, n_j)|$ is the shortest path between $n_i$ and $n_j$ in $G$.

Given $n_i$ and $q$,  eventually, we design our model $F$ score for each $n_j \in N, n_j \not = n_i$ as:
\begin{displaymath}
F(n_j | n_i, q) = I_1 * I_2 = \frac{\sum_{j' \in N(n_j) } e_{ij}}{ |N(n_j)| |SP(n_i, n_j)|}.
\end{displaymath}
For each node of each class, we will compute its $F$ score and recommend top authors to $n_i$.

\input{embedding_model.tex}

%% file: experiments.tex
\section{experiments}
In this part, we will present some results of CQBG-R and CQBG-NE-R and analyze the results. For CQBG-R model, apart from BM25 model, we also use the TF-IDF model to calculate the score of papers. For CQBG-NE-R model, we use various network embedding model to function as the NE model in 
CQBG-NE-R model, For randomwalk based methods, we ran RandomWalk, Node2vec and LINE. For GNN based methods, we ran GCN(Graph Convolutional Network)\cite{kipf2016semi} and PAGNN(position aware graph neural networks)\cite{you2019position}. Because our work is the first academic team worker recommendation work, we do not have some baselines to beat and we can only analyze the results based on the citations or past collaborations between the member of the the team. This part has three subsections: the first is the results of CQBG-R model, the second part is the results of CQBG-NE-R model and the third part is the analysis and discussion on the experiments result.

For the experiment part, we conduct three queries: for each query, the role of the researcher is different and the the interest and tasks for the query are also different. For the three query, the role of researcher for the first query is \textbf{assistant professor} and the researcher is \textbf{Jay Lee}. The research interest is \textbf{Data mining} and the task is \textbf{heterogeneous network}. The role of researcher for the second query is \textbf{prime professor} and the researcher is \textbf{Gamal Fahmy}. The research interest of query 2 is \textbf{computer security} and the task is \textbf{information security}. The role of researcher for the third query is student and the researcher is \textbf{David Kuilman}. The research interest for query 3 is \textbf{computer networks} and the task is \textbf{network algorithm}
\subsection{Results of CQBG-R model}
For CQBG-R model, we ran two different algorithms: TF-IDF and BM25 to calculate the scores of the papers of filtered candidates.
\subsubsection{TF-IDF score calculating result}
For TF-IDF score calculating result in Table2, the first two rows are the top@1, top@2recommending academic team for query 1. The middle two rows are the top@1, top@2 recommending academic team for query 2. The last two rows are the top@1, top@2 recommending academic team for query 3. It is all the same for the remaining model results.
\begin{table}[H]
	\centering
	\caption{TF-IDF score calculating result}
	\begin{tabular}{ccc}
		\toprule  
		Priority &Top@1 &Top@2  \\ 
		\midrule  
		Assistant professor&Terry Coppock&Riichiro Mizoguchi\\
		Student&Haitao Sun&Mathew Scott\\
		\midrule  
		Prime professor&Vijayalakshmi Atluri&Anil K. Jain\\
		Student&Pierre-Alain Fouque&Michael K. Reiter \\
		\midrule  
		Prime professor&Andreas F. Molisch&Chuang Lin\\
		Assistant professor&Subhabrata Sen&Naixue Xiong\\
		\bottomrule  
	\end{tabular}
\end{table}
After evaluation, we find that using TF-IDF to calculate the score of papers is not so accurate. For query 2, we observe closely the paper selected for constructing CQBG and find that this paper is not related closely to the querying task.

\subsubsection{BM25 score calculating result}
the parameters for BM25 score calculating function is set as $k=1.5$ and $b=0.75$. Using BM25 as papers' score function performs better than TF-IDF such as query3 we find that the two professors recommended have published papers related to it.
\begin{table}[H]
	\centering
	\caption{BM25 score calculating result}
	\begin{tabular}{ccc}
		\toprule  
		Priority &Top@1 &Top@2  \\ 
		\midrule  
		Assistant professor&Shih-Chung Kang&Shang-Hsien Hsieh\\
		Student&Jon Wicks&Haoyun Wu\\
		\midrule  
		Prime professor&Min Wu&Allen Roginsky\\
		Student&Vern Paxson&Roberto Di Pietro \\
		\midrule  
		Prime professor&Robert Schober&C. Siva Ram Murthy\\
		Assistant professor&Bo Sun&Ben Y. Zhao\\
		\bottomrule  
	\end{tabular}
\end{table}

\subsubsection{CQBG-PAGNN-R model result}
This model is designed to test the effect of the position on the embedding result. The results in Table 8 show that after adding the position information, the whole framework will recommend some team members, who are close to the querying researcher, which are not what we want.  
\begin{table}
	\centering
	\caption{CQBG-PAGNN-R model result}
\begin{tabular}{ccc}
		\toprule  
		Priority &Top@1 &Top@2  \\ 
		\midrule  
		Assistant professor&Ram D. Sriram&Qi Hao\\
		Student&Kossi P. Adzakpa&Kondo H. Adjallah\\
		\midrule  
		Prime professor&David Wagner&Pierangela Samarati\\
		Student&Christopher Kruegel&Yang Xiao \\
		\midrule  
		Prime professor&Xuemin Shen&Serge Fdida\\
		Assistant professor&Ying-Dar Lin&Song Chong\\
		\bottomrule  
	\end{tabular}
\end{table}

%% file: conclusion.tex
\section{Conclusion}

To the best of our knowledge, we are the first to focus on academic team recommendation, compared to previous person-based recommendation. This problem is challenging as it involves different roles and different types of data. First, it usually is difficult to gain the real academic level of each author and determine the role they can be. Second, the input of this problem includes a text query, which constrains that the recommended author have to have related experiences with this query.